\documentclass{pasa}

\usepackage{graphicx}
\usepackage{amsmath}	
\usepackage{amssymb}	

\usepackage{aas_macros}
\usepackage{hyperref} 
\hypersetup{colorlinks,citecolor=blue,linkcolor=blue,urlcolor=blue}

\usepackage[normalem]{ulem}

\title[Bispectrum with MWA Phase II]{Gridded and direct Epoch of Reionisation bispectrum estimates using the Murchison Widefield Array}

\author[Trott et al.]{Cathryn M. Trott$^{1,2}$\thanks{cathryn.trott@curtin.edu.au}, Catherine A. Watkinson$^3$, Christopher H. Jordan$^{1,2}$, Shintaro Yoshiura$^4$, Suman Majumdar$^{5,3}$, N.~Barry$^{11,2}$,
R.~Byrne$^{8}$,
B.~J.~Hazelton$^{8,12}$,
K.~Hasegawa$^{4}$,
R.~Joseph$^{1,2}$,
T.~Kaneuji$^{4}$,
K.~Kubota$^{4}$,
W.~Li$^{13}$,
J.~Line$^{1,2}$,
C.~Lynch$^{1,2}$,
B.~McKinley$^{1,2}$,
D.~A.~Mitchell$^{10}$,
M.~F.~Morales$^{8,2}$,
S.~Murray$^{1,2,6}$,
B.~Pindor$^{11,2}$,
J.~C.~Pober$^{13}$,
M.~Rahimi$^{11}$,
J.~Riding$^{11}$,
K.~Takahashi$^{4}$,
S.~J.~Tingay$^{1}$,
R.~B.~Wayth$^{1,2}$,
R.~L.~Webster$^{11,2}$,
M.~Wilensky$^{8}$,
J.~S.~B.~Wyithe$^{11,2}$,
Q.~Zheng$^{14}$,
David~Emrich$^1$, A.~P.~Beardsley$^6$, T.~Booler$^1$, B.~Crosse$^1$, T.~M.~O.~Franzen$^{12}$, L.~Horsley$^1$, M.~Johnston-Hollitt$^1$, D.~L.~Kaplan$^7$, 
D.~Kenney$^1$, D.~Pallot$^{9}$, G.~Sleap$^1$, K.~Steele$^1$, M.~Walker$^1$, A.~Williams$^1$, C.~Wu$^{9}$,

\affil{$^1$International Centre for Radio Astronomy Research (ICRAR), Curtin University, Bentley WA, Australia}
\affil{$^2$ARC Centre of Excellence for All Sky Astrophysics in 3 Dimensions (ASTRO 3D), Australia}
\affil{$^3$Department of Physics, Blackett Laboratory, Imperial College, London SW7 2AZ, United Kingdom}
\affil{$^4$Kumamoto University, Japan}
\affil{$^5$Centre of Astronomy, Indian Institute of Technology Indore, Simrol, Indore 453552, India}
\affil{$^6$School of Earth and Space Exploration, Arizona State University, Tempe, AZ 85287, USA}
\affil{$^7$Department of Physics, University of Wisconsin--Milwaukee, Milwaukee, WI 53201, USA}
\affil{$^8$Department of Physics, University of Washington, Seattle, WA 98195, USA}
\affil{$^{9}$International Centre for Radio Astronomy Research (ICRAR), University of Western Australia, Crawley, WA 6009, Australia}
\affil{$^{10}$CSIRO Astronomy \& Space Science, Australia Telescope National Facility, P.O. Box 76, Epping, NSW 1710, Australia}
\affil{$^{11}$School of Physics, The University of Melbourne, Parkville, VIC 3010, Australia}
\affil{$^{12}$University of Washington, eScience Institute, Seattle, WA 98195, USA}
\affil{$^{13}$Brown University, Department of Physics, Providence, RI 02912, USA}
\affil{$^{14}$Shanghai Astronomical Observatory, China}
}

\jid{PASA}
\doi{10.1017/pas.\the\year.xxx}
\jyear{\the\year}

\hypersetup{draft}

\begin{document}

\begin{frontmatter}
\maketitle

\begin{abstract}
We apply two methods to estimate the 21~cm bispectrum from data taken within the Epoch of Reionisation (EoR) project of the Murchison Widefield Array (MWA). Using data acquired with the Phase II compact array allows a direct bispectrum estimate to be undertaken on the multiple redundantly-spaced triangles of antenna tiles, as well as an estimate based on data gridded to the $uv$-plane. The direct and gridded bispectrum estimators are applied to 21 hours of high-band (167--197~MHz; $z$=6.2--7.5) data from the 2016 and 2017 observing seasons. Analytic predictions for the bispectrum bias and variance for point source foregrounds are derived. We compare the output of these approaches, the foreground contribution to the signal, and future prospects for measuring the bispectra with redundant and non-redundant arrays. We find that some triangle configurations yield bispectrum estimates that are consistent with the expected noise level after 10 hours, while equilateral configurations are strongly foreground-dominated. Careful choice of triangle configurations may be made to reduce foreground bias that hinders power spectrum estimators, and the 21~cm bispectrum may be accessible in less time than the 21~cm power spectrum for some wave modes, with detections in hundreds of hours.
\end{abstract}

\end{frontmatter}

\section{INTRODUCTION }
\label{sec:intro}
Exploration of the growth of structure in the first billion years of the Universe is a key observational driver for many experiments. One tracer of the conditions within the early Universe is the 21~cm spectral line of neutral hydrogen, which encodes in its brightness temperature distribution details of the radiation field and gas properties in the intergalactic medium permeating the cosmos \citep{furlanetto06,pritchard08}. Redshifted to low frequencies, the 21~cm line is accessible with radio telescopes ($\nu<300$~MHz), including current and future instruments. These include the Murchison Widefield Array, MWA{\footnote[1]{http://www.mwatelescope.org}} \citep{bowman13_mwascience,tingay13_mwasystem,jacobs16}; the Precision Array for Probing the Epoch of Reionization, PAPER{\footnote[2]{http://eor.berkeley.edu}} \citep{parsons10}; the LOw Frequency ARray, LOFAR{\footnote[3]{http://www.lofar.org}} \citep{vanhaarlem13,patil16}; the Long Wavelength Array, LWA{\footnote[4]{http://lwa.unm.edu}} \citep{ellingson09}, and the future HERA \citep{deboer16} and SKA-Low \citep{koopmans15}.

The weakness of the signal, combined with the expectation that most of its information content is contained in the second moment \citep{wyithe07}, which is uncorrelated across spatial {Fourier wave mode}, motivates the use of the power spectrum as a statistical tool for detecting and characterising the cosmological signal. Despite the ease with which the power spectrum can be computed from radio interferometric data, the presence of strong, spectrally-structured residual foreground sources \citep{trott12,datta10,vedantham12,thyagarajan15a}, complex instrumentation \citep{trottwayth2016}, and imperfect calibration \citep{patil14,barry16}, yield power spectra that are dominated by systematics. Thus far, a detection of signal from the Epoch of Reionisation has not been achieved \citep{patil16,beardsley16,trottchips2016,cheng18}. These systematics, combined with the expectation that non-Gaussian information can be extracted usefully from cosmological data, lead the discussion for other statistics. The bispectrum, as a measure of signal non-Gaussianity, is one such statistic that contains cosmologically relevant information \citep{bharadwaj05,majumdar18,watkinson18}, while being relatively straightforward to compute with interferometric data \citep{shimabukuro17}.

The bispectrum is the Fourier Transform of the three-point correlation function, and extracts {higher-order} correlations between different spatial scales. Its spatial and redshift evolution can be used to place different constraints on the underlying processes that set the 21~cm brightness temperature, and therefore it provides complementary information to the power spectrum. In an early paper exploring the use of the bispectrum for a model EoR signal, and radio interferometers, \citet{bharadwaj05} demonstrated that a strong non-Gaussian signal is produced by the presence of ionized regions, and discussed the behaviour of the power spectrum and bispectrum signals as a function of frequency channel separation, although they only consider non-Gaussianity due to the ionisation field modelled as non-overlapping randomly placed spherical ionised regions. Some recent work has explored the combination of bispectrum with other tracers (CII spectral features) to extract clean cosmological information \citep{beane18}. {The bispectrum has also been used in the single-frequency (angular) case in the CMB community, where non-Gaussianities can be contaminated by structured foregrounds \citep{jung18}.}

\citet{majumdar18} explore the ability of the bispectrum to discriminate fluctuations in the matter density distribution from those of the hydrogen neutral fraction, reporting that for some triangle configurations the sign of the bispectrum is a marker for which of these processes is dominating the bispectrum. They show output bispectra for equilateral and isosceles configurations over a range of wavemodes and redshifts, including parameters of relevance to current low-frequency 21~cm experiments ($z<9, 0.1 < k < 1.0$). 
For modes relevant to the MWA, the bispectrum amplitude fluctuates in sign with wavenumber and triangle geometry (stretched $\rightarrow$ equilateral $\rightarrow$ squeezed) with a range spanning $10^3 - 10^9$~mK$^3 h^{-6}$ Mpc$^6$. This range of potential signs and amplitudes in measurable modes and redshifts, motivates us to study this signal in MWA data.

\citet{watkinson18} provide a useful tool for visualising the correspondence of real-space structures and bispectrum. They highlight that equilateral $k$-vector configurations probe above-average signal concentrated in filaments with a circular cross section (their Figure 1). Stretched (flattened) $k$-vector triangle configurations (with one $k$-mode larger than the other two), by extension, probe above-average signal concentrated in filaments with ellipsoidal cross sections (at the extreme these filaments tend towards planes). 
Finally, squeezed $k$-vector triangle configurations (with one $k$-mode smaller than the other two) correspond to a modulation of a large-scale mode over small-scale plane-wave concentrations of above-average signal, and therefore measure the correlation of the small-scale power spectrum with large-scale modes.

Notably, they introduce and explore other bispectrum normalisations that are found to be more stable to parameter fluctuations. In this work, we discuss the relative merits of different bispectrum statistics for use with real data in the presence of real systematics.

Crucially, the switch to positive bispectrum at the end of reionisation occurs as we reach regimes/scales at which the concentration of above-average signal drive the non-Gaussianity. This will occur before the EoR (on scales where the density field is the dominant driver of the temperature fluctuations, or, if the spin temperature is not yet saturated during this phase, when heated regions are driving the non-Gaussianity) and towards the end of reionisation (when islands of 21-cm signal drive the non-Gaussianity).

Conversely, a negative-valued bispectrum will be unique to the phase when ionised regions drive the non-Gaussianity. In general, foreground astrophysical processes are not expected to produce a negative bispectrum, {because they are associated with overdensities in the brightness temperature distribution \citep{lewis11,watkinson14a}}. These factors may play a future important role in discriminating real cosmological non-Gaussianity from contaminants.

Despite some work studying the sensitivity of current and future experiments for measuring the bispectrum \citep{shimabukuro17,yoshiura15}, these have used idealised scenarios that omit any residual foreground signal and systematics introduced by the instrument. \citet{bharadwaj05} discuss foreground fitting tools using frequency separation to study the bispectrum over visibility correlations across frequency, but this method breaks down for large field-of-view instruments where the interferometric response affects the foreground smoothness \citep{morales12}. Further, no 21~cm interferometric data has been used to estimate the bispectrum. In this work, we address both of these by presenting bispectrum estimators that can use real datasets, computing the expected impact of foregrounds measured by the instrument, and applying the estimators to 21 hours of MWA EoR data.

\section{MWA Phase II Array}
The Murchison Widefield Array is a 256-tile low-frequency radio interferometer located in the Western Australian desert, on the future site of the Square Kilometre Array (SKA) \citep{tingay13_mwasystem,bowman13_mwascience}. The telescope operates from 80--300~MHz with antennas spread over a 5~km diameter. Its primary science areas include exploration of the Epoch of Reionisation, radio transients, solar and heliospheric studies, study of pulsars and fast transients, and the production of a full-sky low-frequency extragalactic catalogue. In 2016 it underwent an upgrade from 128 to 256 antenna tiles \citep{wayth18}. At any time, 128 of the tiles can be connected to the signal processing system. The array operates in a "compact" configuration, utilising redundant spacings and short baselines for EoR science, or an "extended" configuration, maximising angular resolution and instantaneous $uv$-coverage. The compact configuration is employed in this work.

\begin{figure}
\begin{center}
\includegraphics[width=20pc]{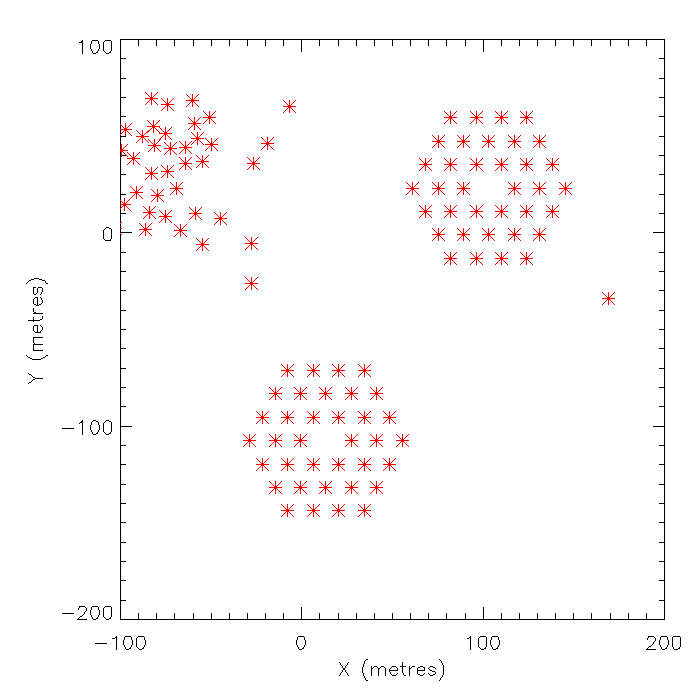}
\caption{Zoomed MWA compact configuration layout showing the two hexagonal subarrays of 36 tiles each, with redundant tile spacings. These short redundant baselines are used in this work to form equilateral and isosceles triangle bispectra with high sensitivity. Some of the longer baseline tiles of the MWA are not shown.}
 \label{fig:arrayhex}
\end{center}
\end{figure}

The compact configuration has a maximum baseline of 500~metres and is optimised for EoR science. Figure \ref{fig:arrayhex} shows the tile layout, including the two 36-tile hexagonal subarrays of redundantly-spaced tiles. The minimum redundant spacing is 14~m.
The primary motivations for the hexagons are two-fold: (1) to increase the sensitivity to angular scales of relevance for the EoR, allowing coherent addition of measurements from redundant baselines, and (2) enabling additional methods for calibrating the array (redundant calibration, Li et al. 2018, Joseph et al. 2018). For the bispectrum, there is an additional advantage of multiple, redundant equilateral triangle baselines being formed from the short spacings. These can be added coherently to study the bispectrum signal on particular scales, and allows for a direct bispectrum measurement (perfectly-defined triangles formed from discrete baselines). These direct bispectrum results can be compared to a more general gridded bispectrum, whereby all baselines formed by an irregularly-spaced array (such as MWA Phase I, or the non-hexagon tiles of Phase II compact) can be gridded onto the Fourier ($uv$-) plane, using a gridding kernel that represents the Fourier response function of the telescope (in this case, the Fourier Transform of the primary beam response to the sky). These estimators will both be explored in this work.

\section{Power spectrum}
We briefly review the power spectrum as the primary estimator for studying the EoR with 21~cm observations. The power spectrum is typically used to describe radio interferometer observations from the EoR, and contains all of the Gaussian-distributed fluctuation information. The power spectrum is the power spectral density of the spatial fluctuations in the 21~cm brightness temperature field. It is used because it encodes the fluctuation variance (where most of the EoR signal is expected to reside), and sums signal from across the observing volume to increase sensitivity. It is defined as:
\begin{equation}
    P(\vec{k}) = \delta_D(\vec{k}-\vec{k}^\prime) \frac{1}{\Omega_V} \langle V^\ast(\vec{k}) V(\vec{k}^\prime) \rangle,
\end{equation}
where $V(\vec{k}) = V(u,v,\eta) = \mathcal{FT} \left(V(u,v,\nu)\right)$ is the measured interferometric visibility (Jansky), Fourier Transformed along frequency ($\nu$) to map frequency to line-of-sight spatial scales (Jy Hz) at a given point in the Fourier ($uv$-) angular plane ($u,v$); $\langle \rangle$ encode an ensemble average over different realisations of the Universe, and the $\delta_D$-function ensures that we are expecting to measure a Gaussian random field where the different modes are uncorrelated\footnote{The mapping from observed to cosmological dimensions is given by:
\begin{eqnarray}
k_\bot &=& \frac{2\pi{|\boldsymbol{u}|}}{D_M(z)},\\
k_\parallel &=& \frac{2\pi{H_0}f_{21}E(z)}{c(1+z)^2}\eta,
\end{eqnarray}
where $D_M$ is the transverse comoving distance, and $f_{21}$ is the rest frequency of the neutral hydrogen emission.}. Further assuming spatial isotropy allows us to average incoherently in spherical shells, where $\vec{k} = k$. $\Omega_V$ provides the volume normalisation, where $\Omega_V = (\rm{BW})\Omega$ is the product of the observing bandwidth and angular field-of-view. Converting from measured to physical units maps Jy$^2$ Hz$^2$ to mK$^2$ $h^{-6}$Mpc$^6$. After volume normalisation this becomes, mK$^2$ $h^{-3}$Mpc$^3$.


\subsection{Power spectra with radio interferometric data}
The power spectrum can be produced naturally with interferometric data. Unlike optical telescopes that produce images of the sky, or single-dish radio telescopes that acquire a single sky power, a radio interferometer visibility (Jy) directly measures Fourier representations of the sky brightness distribution at the projected baseline location ($u=\Delta{x}/\lambda; v=\Delta{y}/\lambda$). In the flat-sky approximation\footnote{This is appropriate for this work where the data used are all from zenith-pointed snapshots, where the $w$-terms are small.}:
\small
\begin{equation}
    V(u,v,\nu) = \displaystyle\int_\Omega A(l,m,\nu) S(l,m,\nu) \exp{(-2\pi{i}(ul+vm))} dldm,
    \label{eqn:vis}
\end{equation}
\normalsize
where $A(l,m,\nu)$ is the instrument primary beam response to the sky at position $(l,m)$ from the phase centre and frequency $\nu$, $S(l,m,\nu)$ is the corresponding sky brightness (Jy/sr, which is proportional to temperature), and the exponential encodes the Fourier kernel. The physical correspondence of sky projected on to the tile locations yields a fixed set of discrete but incomplete Fourier modes to be measured. This incompleteness leads to parts of the Fourier plane where there is no information. The line-of-sight spatial scales are obtained by Fourier Transform of visibilities measured at different frequencies, along frequency to map $\nu$ to $\eta$:
\begin{equation}
    V(\eta(k)) = \mathcal{FT} (V(\nu)) = \frac{\Delta\nu}{N_{\rm ch}} \displaystyle\sum_{j=1}^{N_{\rm ch}} V(\nu)\exp{\left(-\frac{2\pi{i}jk}{N_{\rm ch}}\right)},
\end{equation}
where $N_{\rm ch}$ is the number of spectral channels, $\Delta\nu$ is the spectral resolution, and $j$ and $k$ index frequency and spatial mode (Hz$^{-1}$, or seconds).

The attenuation of the sky due to the primary beam (and general sky finiteness) alters the complete continuous Fourier Transform to a windowed transform, whereby the primary beam response leaks signal into adjacent Fourier modes, as can be seen using the convolution theorem:
\begin{equation}
    V(u,v,\eta) = \tilde{A}(u,v,\eta) \circledast \tilde{S}(u,v,\eta),
    \label{eqn:convolution}
\end{equation}
where the true sky brightness distribution is convolved with the Fourier Transform of the primary beam response. This leakage implies that the visibility measured by a discrete baseline actually contains signal from a region of the Fourier plane, as described by the Fourier beam kernel, $\tilde{A}(u,v,\eta)$.

In general, to compute the power spectrum from a large amount of data, we are motivated by the signal weakness to add the data coherently; i.e., we sum complex visibilities directly that contribute signal to the same point in the Fourier $uv$-plane. To do this, the measurement from each baseline is convolved with the Fourier beam kernel and `gridded' (added with a weight) onto a common two-dimensional plane. Signal will add coherently, while noise adds as the square-root (because the thermal noise is uncorrelated between measurements). The weights for each measurement are also gridded with the kernel onto a similar plane. After addition of all the data, the signal $uv$-plane is divided by the weights to yield the optimal-weighted average signal at each point. The resulting cube resides in $(u,v,\nu)$ space, and can be Fourier Transformed along frequency to obtain a cube in $(u,v,\eta)$ space. The power spectrum can then be formed by squaring and normalising the cube, and averaging incoherently (in power) in spherical shells:
\begin{equation}
    P(|\vec{k}|) = \frac{\displaystyle\sum_{i \in k} V^{\ast}_i(\vec{k})V_i(\vec{k})W_i(\vec{k})}{\displaystyle\sum_{i \in k} W_i(\vec{k})},
\end{equation}
where $W$ are the weights and $|\vec{k}| = |(k_u,k_v,k_\eta)| = \sqrt{k_u^2+k_v^2+k_\eta^2}$.

As an intermediate step, the cylindrically-averaged power spectrum can be formed \citep[e.g.,][]{datta10}:
\begin{equation}
    P(k_\bot,k_\parallel) = \frac{\displaystyle\sum_{i \in k_\bot} V^{\ast}(\vec{k})V(\vec{k})W(\vec{k})}{\displaystyle\sum_{i \in k_\bot} W(\vec{k})},
\end{equation}
and $k_\bot = \sqrt{k_u^2+k_v^2}$, $k_\parallel=k_\eta$. This is a useful estimator for discriminating contaminating foregrounds (continuum sources with power concentrated at small $k_\parallel$) from 21~cm signal. Herein we will refer to this power spectrum, and its bispectrum analog, as the `gridded power spectrum' and `gridded bispectrum', respectively.

Alternatively, one can take the baselines themselves, and their visibilities measured along frequency, and take the Fourier Transform directly along the frequency axis. This `delay spectrum' approach is utilised by some experiments with short baselines \citep{parsons12,ali15,thyagarajan15a}, both to increase sensitivity when there are redundant spacings, and to work as a diagnostic. The frequency and $\eta$ axes are not parallel, except at zero-length baseline. Because an interferometer is formed instantaneously from antennas with a fixed spatial offset, the baseline length in Fourier space (e.g., $u$) evolves with frequency as $u=\Delta{x}\nu/c$, and this evolution is therefore increased for larger bandwidths and for longer baselines. For the short spacings of interest to the EoR, the correspondence is good, and the delay transform can be used to mimic the direct $k_\parallel$ transform of gridded data \citep[see Figure 1 of][for a visual explanation]{morales12}. In general, `imaging' arrays with many non-redundant spacings are suited to gridded power spectra, whereas redundant arrays, with a lesser number of multiply-sampled modes, are suited to delay power spectra. For the Phase II compact MWA, the two hexagonal subarrays have these short-spaced redundant baselines, and the `delay power spectrum' and its bispectrum analog can also be used effectively. In general, we would not suggest use of the delay spectrum to undertake EoR science, because of the limitations discussed, but it is the appropriate analogue for the direct bispectrum estimator, and is therefore pertinent for the normalised bispectrum analysis.

\section{Bispectrum}
The bispectrum is the Fourier Transform of the three-point correlation function. Akin to the two-point correlation function (the Fourier dual of which is the power spectrum), the three-point correlation function measures the excess signal over that of a Gaussian random field distribution measured at three spatial locations, averaged over the volume. For a field with Fourier Transform denoted by $\Delta(\vec{k})$, the bispectrum is formed over closed triangles of $k$ vectors in Fourier space:
\begin{equation}
    \langle \Delta(\vec{k}_1)\Delta(\vec{k}_2)\Delta(\vec{k}_3) \rangle  = \delta_D(\vec{k}_1,\vec{k}_2,\vec{k}_3) \rm{B}(\vec{k}_1,\vec{k}_2,\vec{k}_3).
\end{equation}
Here the $\delta_D$-function ensures closure in Fourier space. It has units of mK$^3$ $h^{-6}$Mpc$^6$ after volume normalisation.
The bispectrum is often applied to matter density fields, where $\Delta(\vec{k})$ is the Fourier Transform of matter overdensity, $\delta(\vec{x}) = \frac{\rho(\vec{x})}{\overline\rho} - 1$. In radio interferometric measurements, the coherence of the wavefront (the visibilities obtained by cross-correlating voltages from individual antennas) represents the Fourier Transform of the sky brightness temperature distribution, measured in Jansky.

As discussed earlier, this bispectrum estimator can be unstable, with cosmological simulations showing rapid fluctuations between positive and negative values as non-Gaussianity becomes negligible but the amplitude is still large. As such, \citet{watkinson18} suggest the normalised bispectrum as a more stable statistic:
\begin{equation}
    \mathcal{B}(\vec{k}_1,\vec{k}_2,\vec{k}_3) = \frac{\rm{B}(\vec{k}_1,\vec{k}_2,\vec{k}_3)\sqrt{k_1k_2k_3}}{\sqrt{P(\vec{k}_1)P(\vec{k}_2)P(\vec{k}_3)}},
\end{equation}
where $P(|\vec{k}|)$ is the three-dimensional power spectrum, which describes the volume-normalised variance on a given spatial scale, and is the Fourier Transform of the two-point correlation function \citep{eggemeier2017,brillinger67}. 
This normalisation {isolates the contribution from the non-Gaussianity to the bispectrum, by normalising out the amplitude part of the statistic}.
It is akin to normalising the 3rd central moment by $\sigma^3$ to calculate the skewness.

\subsection{Bispectrum with radio interferometric data}
Because the bispectrum is formed from the triple product of a triangle of wavespace measurements, it can be formed directly through the product of three interferometric visibilities. In the limit where the array has perfect (complete) $uv$-sampling, individual measurements of signal on triangles of baselines can be multiplied to form the bispectrum estimate. In the more general case, where an interferometer has instantaneously incomplete, but well-sampled baselines, there are two options for extracting the triangles of signal measurements: direct (via multiplication of measurements from three tiles forming a triangle of baselines), or gridded, where each $uv$-measurement is gridded onto the $uv$-plane (with its corresponding Fourier beam kernel and weights), and the final bispectra are computed from the fully-integrated and gridded data.

Direct bispectrum estimators can be applied to specific triangles according to the array layout, but these are usually unique, with irregular configurations (all three internal angles are distinct), leading to difficult cosmological interpretation and poor sensitivity. These issues arise for imaging-like arrays with pseudo-random layouts, but are alleviated for redundant arrays, where regular triangles (isosceles and equilateral) exist and are instantaneously available in many copies in the array. These features make interpretation more straight-forward and increase sensitivity to these bispectrum modes.

Gridded bispectrum estimators can be applied to any array, yield improved sensitivity by coherent gridding of data and may allow for a wider range of triangles to be probed. Nonetheless, they suffer from the increased difficulty of extracting robust estimates that correctly account for the correlation of data in $uv$-space.

With the benefit of having a redundant array, we will apply both sets of estimators to our data.

\section{The Gridded Estimator}\label{sec:gestimator}
Each measured visibility encodes information about a small range of Fourier modes of the sky brightness distribution. Although each baseline is usually reported as a single number representing the antenna separation measured between antenna centres, the baselines actually measure a range of separations when accounting for the actual physical size\footnote{Due to the physical size of the collecting antenna element, some parts of the antenna have a smaller effective baseline length (closer to other antenna), and some have a longer (further from the other antenna).}. This translates to a range of Fourier modes being measured by a given baseline, and is equivalent to the statement that a finite primary beam response to the sky mixes Fourier modes through spectral leakage (effectively a taper on the continuous Fourier transform). Thus, when measurements from different baselines are combined coherently (with phase information) onto a $uv$-plane, they can be gridded with a kernel that is the Fourier Transform of the primary beam response to the sky. Such a gridding kernel captures the degree of spectral leakage introduced by the antenna response, and means that baselines of similar length and orientation have some shared information. The gridding kernel is represented by $\tilde{A}$ in Equation \ref{eqn:convolution}. With a single defined visibility phase centre, all visibility measurements can be added with this beam kernel onto a single plane (for each frequency channel), along with their associated weights, to form a coherently-averaged estimate for the Fourier representation of the sky brightness temperature:
\begin{equation}
    \hat{V}_{uv} = \frac{\displaystyle\sum_i V(u_i,v_i)\tilde{A}(u_i,v_i)W(u_i,v_i)}{\displaystyle\sum_i \tilde{A}(u_i,v_i)W(u_i,v_i)},
\end{equation}
where $i$ indexes measurement, and $W$ is the weight associated with each. The bispectrum is then estimated as the sum over the beam-weighted gridded visibilities:
\begin{equation}
    \hat{B}_{123} = \frac{\displaystyle\sum_{j \in \tilde{A}} \hat{V}_{j1}\hat{V}_{j2}\hat{V}_{j3} W_{j1}W_{j2}W_{j3}}{\displaystyle\sum_j W_{j1}W_{j2}W_{j3}},
\end{equation}
where
\begin{equation}
    W_{1j} = W_{1}\tilde{A}_{1j},
\end{equation}
are the beam-gridded measurement weights.

\subsection{Gridded Estimator noise}
The gridded bispectrum estimator is formed from coherent addition of visibilities over \textit{all} observations. As such, if a given visibility has thermal noise level $\sigma_{\rm therm}$ (Jy Hz)\footnote{$\sigma_{\rm therm} = \frac{2kT}{\lambda^2}\Omega\frac{\Delta{\nu}}{\sqrt{\rm{BW}\Delta{t}}}$ for bandwidth BW, spectral resolution $\Delta\nu$ and observation time interval $\Delta{t}$}, the uncertainty on the bispectrum is:
\begin{equation}
    \Delta\hat{B}_{\rm TOT} = \frac{\sqrt{3}\sigma_{\rm therm}^3}{\sqrt{\displaystyle\sum_{\tilde{A}} W_{1}W_{2}W_{3}}},
    \label{eqn:errorbispec}
\end{equation}
where the denominator is the sum over the gridding kernel of the weights triplets.

The uncertainty on the normalised bispectrum is then:
\begin{equation}
    \Delta\hat{\mathcal{B}}_{123, {\rm TOT}} = \mathcal{B} \sqrt{\frac{\Delta{B}^2}{B^2} + \frac{\Delta{P}_1^2}{4P_1^2} + \frac{\Delta{P}_2^2}{4P_2^2} + \frac{\Delta{P}_3^2}{4P_3^2}},
    \label{eqn:error}
\end{equation}
{where the uncertainties can contain both thermal noise and noise-like uncertainty from residual foregrounds.}

For 300 2-minute observations, and 24 triangles per 28~m baseline triad group, the expected {thermal} noise level for a complete dataset is:
\begin{equation}
    \Delta{B} = 4.2 \times 10^{10} {\rm mK}^3 h^{-6} {\rm Mpc}^6.
\end{equation}
{The presence of residual foregrounds will be studied in Section \ref{sec:fg_var}}.

\section{The Direct Estimator}\label{sec:destimator}
As an alternate approach to the gridded estimator, visibilities are Fourier-transformed along the frequency direction to compute the delay transform, and closed bispectrum triangles formed from the closed redundant triads of antennas. This approach does not use the primary beam, and ignores the local spatial correlations generated by the primary beam spatial taper. It also transforms along a dimension that changes angle with respect to $k_\parallel$ as a function of baseline length, but approximates a $k_\parallel$ Fourier Transform for small $u$ (small angle).

The bispectrum for a given observation is the weighted average over all triads:
\begin{equation}
    \hat{B}_{123} = \frac{(\displaystyle\sum_{i} V_{1i}W_{1i})(\displaystyle\sum_{i} V_{2i}W_{2i})(\displaystyle\sum_{i} V_{3i}W_{3i})}{(\displaystyle\sum_{i} W_{1i})(\displaystyle\sum_{i} W_{2i})(\displaystyle\sum_{i} W_{3i})},
\end{equation}
where $i$ indexes over redundant triangles (triads). The final bispectrum estimate then performs a weighted average over observations, such that:
\begin{equation}
    \hat{B}_{123,{\rm TOT}} = \frac{\displaystyle\sum_j \hat{B}_{123,j} W_j}{\displaystyle\sum_j W_j},
\end{equation}
where
\begin{equation}
    W_j = (\displaystyle\sum_{i} W_{1i})(\displaystyle\sum_{i} W_{2i})(\displaystyle\sum_{i} W_{3i}).
\end{equation}

\subsection{Direct Estimator noise}
The direct bispectrum estimator is formed from coherent addition of baseline triplets for a given observation, which are then averaged with relative weights to the final estimate. As such, if a given visibility has thermal noise level $\sigma_{\rm therm}$ (Jy Hz), the uncertainty on the bispectrum is:
\begin{equation}
    \Delta\hat{B}_{\rm TOT} = \frac{\sqrt{3}\sigma_{\rm therm}^3}{\sqrt{\displaystyle\sum_j W_j}}.
\end{equation}

The uncertainty on the normalised bispectrum is then given by the same expression as for the Gridded Estimator (Equation \ref{eqn:error}).

For 300 observations, and 24 triangles per 28~m baseline triad group, the expected noise level is:
\begin{equation}
    \Delta{B} = 7.1 \times 10^{11} {\rm mK}^3 h^{-6} {\rm Mpc}^6.
\end{equation}

\section{Triangles considered for estimation}
Unlike bispectrum estimates that can be obtained from Phase I data, where the array is in an imaging configuration with no redundant triangles, we aim to take advantage of the 72 redundant tiles in the hexagonal sub-arrays, afforded by the Phase II layout. This allows for both direct and gridded bispectrum estimators to be applied to matched observations with matched data calibration.

The most numerous (highest sensitivity) groups of redundant triangles are the \textit{angularly}-equilateral configurations of the 14~m and 28~m baselines (48 and 24 sets, respectively). For these triangles, the equilateral configurations exist only for the $\eta=0$ ($k_\parallel=0$) line-of-sight mode. Other configurations of these closed angular triangles are isosceles or irregular triangles, depending on the $\eta$ values chosen, however the closed triangle requirement of the bispectrum demands that:
\begin{equation}
    \eta_1 + \eta_2 + \eta_3 = 0,
\end{equation}
in addition to the angular components of the vectors summing to zero (as is enforced by choosing the closed triangle baselines).

\begin{figure*}
\begin{center}
\includegraphics[width=30pc]{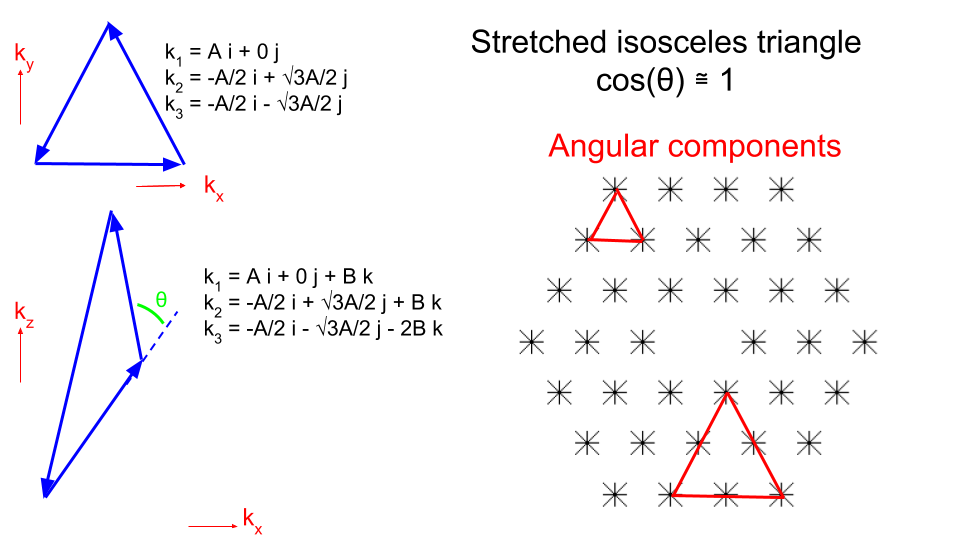}
\caption{Schematic of how a stretched isosceles triangle configuration is extracted from redundant \textit{angularly-equilateral} triangle baselines of the MWA Phase II hexagons.}
 \label{fig:bispec_vec1}
\end{center}
\end{figure*}

\begin{figure*}
\begin{center}
\includegraphics[width=40pc]{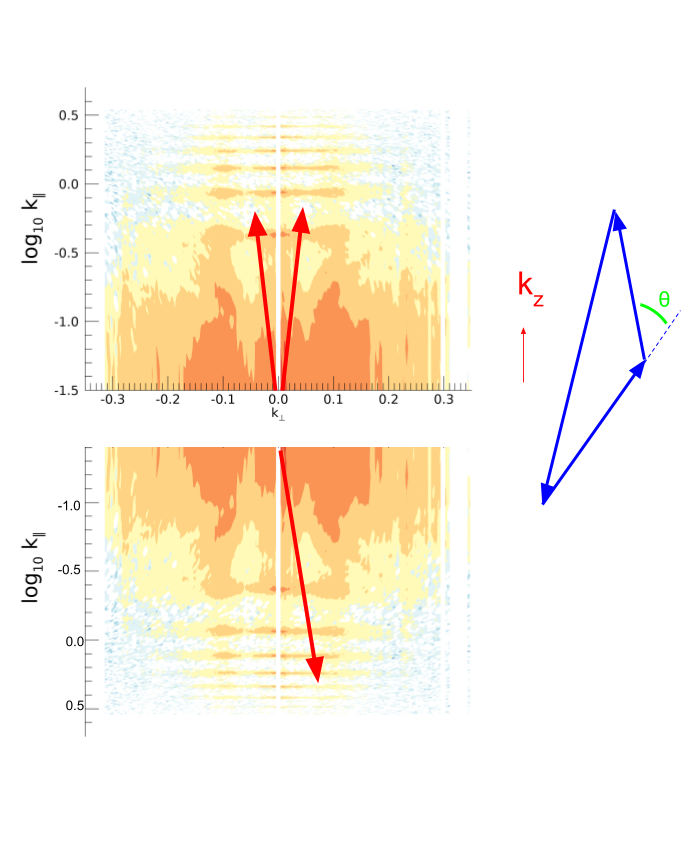}
\caption{Schematic of how isosceles triangle vectors are extracted, overlaid on a power spectrum. We aim to choose triangles with vectors that reside in noise-like regions of the delay spectrum.}
 \label{fig:hex_vec1}
\end{center}
\end{figure*}

For comparison with theoretical predictions, we will focus on equilateral and isosceles triangles. The 14~m and 28~m baselines are very short, corresponding to cosmological scales of $k_\bot \simeq 0.01h$Mpc$^{-1}$ at $z=9$. Thus, although the equilateral configuration is cosmologically relevant and the easiest to interpret, these modes are expected to be heavily foreground dominated (i.e., they correspond to the line-of-sight DC mode, and the large angular scales of diffuse and point source foreground emission). We consider them for completeness, but will show them to be cosmologically irrelevant from an observational perspective when computed this way. These same angularly-equilateral triangle configurations will, however, be used to form relevant isosceles configurations with $\eta_1=\eta_2$ and $\eta_3 = -2\eta_1$. Given that we aim to sample modes where foregrounds are not dominant in our power spectra, these isosceles configurations form `stretched' \citep[also referred to `flattened' in][]{watkinson18} configurations ($k_\parallel >> k_\bot$). Figure \ref{fig:bispec_vec1} shows how the stretched isosceles configurations are extracted from the data with a redundant baseline triad.
Figure \ref{fig:hex_vec1} then shows schematically the approximate vectors for two of the four isosceles configurations considered here.

\section{Observations}
The direct and gridded estimators are applied to 21.0 hours of Phase II high-band zenith-pointed data, comprising 10.7 hours (320 observations) on the EoR0 field (RA$=0$~h, Dec.$=-27$~deg.) and 10.3 hours (309 observations) on the EoR1 field (RA$=4$~h, Dec.$=-30$~deg.). We observe 30.72~MHz in 384 contiguous 80~kHz channels, with a base frequency of 167.035~MHz. Approximately 15\% of the observations were obtained from drift-scan data, where the telescope remains pointed at zenith for many hours and the sky drifts through. For consistency with the drift-n-shift data, we chose drift scan data observed with the field phase centres within 3 degrees of zenith. The data were observed over five weeks from 2016 October 15 to November 28, and one week in 2017 July. Because the delay spectrum is used as part of the power spectrum estimator for the direct bispectrum, each observation was individually inspected for poor calibration or data quality, and bad observations excised from the dataset. The excised observations comprised $\sim$5 percent of the dataset, and primarily were due to poor calibration solutions over sets of data contiguous in time due to poor instrument conditions (e.g., many flagged tiles or spectral channels).

The 2-minute observations were each calibrated through the MWA Real Time System (RTS; Mitchell et al. 2008), as is routinely performed for MWA EoR data, and one thousand of the brightest (apparent) sources peeled from the dataset \citep{jacobs16}. These 629 calibrated and peeled observations were used for bispectrum estimation.

\section{Results}\label{sec:results}
We begin by reporting the bispectrum estimates for the two methods and fields, and then report the normalised bispectra, which incorporate the power spectrum estimates. Table \ref{table:bispectra} shows the bispectrum estimates and their one sigma uncertainties (thermal noise) for the direct and gridded estimators, both observing fields and for different triangle configurations.
\begin{table*}
\centering
\begin{tabular}{|c||c|c||c|c|}
\hline 
Triangles & Direct & Gridded & Type & Faint Galaxy \\
\hline
{\bf EoR0} & ($\times 10^{12}$ mK$^3$ Mpc$^6$) & ($\times 10^{12}$ mK$^3$ Mpc$^6$) & {\bf 14m} & {(mK$^3$ Mpc$^6$)}\\
\hline
$k_1=k_2=k_3=0.007$ & $1.3e9 \pm 7.8$ & $4.3e8 \pm 0.2$ & Equilateral & \\
$k_1=0.2,k_2=k_3=0.1$ & $-1071.2 \pm 7.8$ & $-1.6e4 \pm 0.2$ & Isosceles & 4.4 $\times$ 10$^9$\\
$k_1=0.4,k_2=k_3=0.2$ & $-7571 \pm 7.8$ & $8.9e4 \pm 0.2$ & Isosceles & $-$2.7 $\times$ 10$^7$\\
$k_1=0.6,k_2=k_3=0.3$ & $27250 \pm 7.8$ & $-1078 \pm 0.2$ & Isosceles & $-$3.6 $\times$ 10$^6$\\
$k_1=1.0,k_2=k_3=0.5$ & $47.0 \pm 7.8$ & $22.0 \pm 0.2$ & Isosceles & 5.8 $\times$ 10$^4$ \\
\hline
{\bf EoR0} & & & {\bf 28m}\\
\hline
$k_1=k_2=k_3=0.014$ & $-1.3e7 \pm 22.2$ & $6.9e8 \pm 0.3$ & Equilateral & \\
$k_1=0.2,k_2=k_3=0.1$ & $120.8 \pm 22.2$ & $9582 \pm 0.3$ & Isosceles &\\
$k_1=0.4,k_2=k_3=0.2$ & $-2010 \pm 22.2$ & $-84.0 \pm 0.3$ & Isosceles &\\
$k_1=0.6,k_2=k_3=0.3$ & $943.2 \pm 22.2$ & $65.1 \pm 0.3$ & Isosceles &\\
$k_1=1.0,k_2=k_3=0.5$ & {$\bf 13.7 \pm 22.2$} & $88.2 \pm 0.3$ & Isosceles & \\
\hline \hline
{\bf EoR1} & ($\times 10^{12}$ mK$^3$ Mpc$^6$) & ($\times 10^{12}$ mK$^3$ Mpc$^6$) &  {\bf 14m} & Bright Galaxy \\
\hline
$k_1=k_2=k_3=0.007$ & $-9.9e6 \pm 2.3$ & $2.0e10 \pm 0.3$ & Equilateral & \\
$k_1=0.2,k_2=k_3=0.1$ & $-21.5 \pm 2.3$ & $1.9e4 \pm 0.3$ & Isosceles & 4.4 $\times$ 10$^9$\\
$k_1=0.4,k_2=k_3=0.2$ & $978.9 \pm 2.3$ & $-4.0e8 \pm 0.3$ & Isosceles & $-$2.9 $\times$ 10$^7$\\
$k_1=0.6,k_2=k_3=0.3$ & $1546.4 \pm 2.3$ & $-25.4 \pm 0.3$ & Isosceles & $-$8.4 $\times$ 10$^5$\\
$k_1=1.0,k_2=k_3=0.5$ & {$\bf -2.0 \pm 2.3$} & {$\bf 0.4 \pm 0.3$} & Isosceles & 1.5 $\times$ 10$^5$\\
\hline
{\bf EoR1} & & &  {\bf 28m} \\
\hline
$k_1=k_2=k_3=0.014$ & $3.6e5 \pm 6.7$ & $-1.2e8 \pm 0.5$ & Equilateral & \\
$k_1=0.2,k_2=k_3=0.1$ & {$\bf 2.7 \pm 6.7$} & $-1530 \pm 0.5$ & Isosceles & \\
$k_1=0.4,k_2=k_3=0.2$ & {$\bf 1.7 \pm 6.7$} & $203.3 \pm 0.5$ & Isosceles &\\
$k_1=0.6,k_2=k_3=0.3$ & $-229.4 \pm 6.7$ & $-12.5 \pm 0.5$ & Isosceles &\\
$k_1=1.0,k_2=k_3=0.5$ & $37.1 \pm 6.7$ & $474.3 \pm 0.5$ & Isosceles &\\
\hline \hline
\end{tabular}
\caption{Bispectrum estimates and one sigma uncertainties for the direct and gridded bispectra for each observing field and triangle type. Bold-faced values indicate bispectrum estimates that are consistent with thermal noise. The right-hand column lists expected bispectrum values from simulation for faint and bright galaxies driving reionisation. $k$ modes are comoving and measured in $h$~Mpc$^{-1}$.}\label{table:bispectra}
\end{table*}
Bold-faced results indicate bispectrum estimates that are consistent with thermal noise. These tend to be those that are extremely stretched isosceles configurations ($\cos{\theta} \sim 1$), with estimates that sit well outside the primary foreground contamination parts of parameter space. Conversely, the equilateral triangle configurations that use the $k_\parallel=0$ mode exclusively show extremely large detections. There is no suggestion that these are 21~cm cosmological bispectrum detections, but rather are foreground contaminants. This will be explored more fully in Section \ref{sec:fg}. Note also that the thermal noise levels reported here are a factor of a few larger than the theoretical expectation derived in Sections \ref{sec:gestimator}--\ref{sec:destimator}. This is due to the fraction of data with weights that are less than unity, indicating flagged baselines and spectral channels.

Also listed in Table \ref{table:bispectra} is the expected bispectrum values from simulations that assume either bright or faint galaxies drive reionization \citep{greig17}. The largest amplitudes are for the smallest $k$-modes, which also tend to be more foreground dominated.

\begin{figure*}
\begin{center}
\includegraphics[width=40pc]{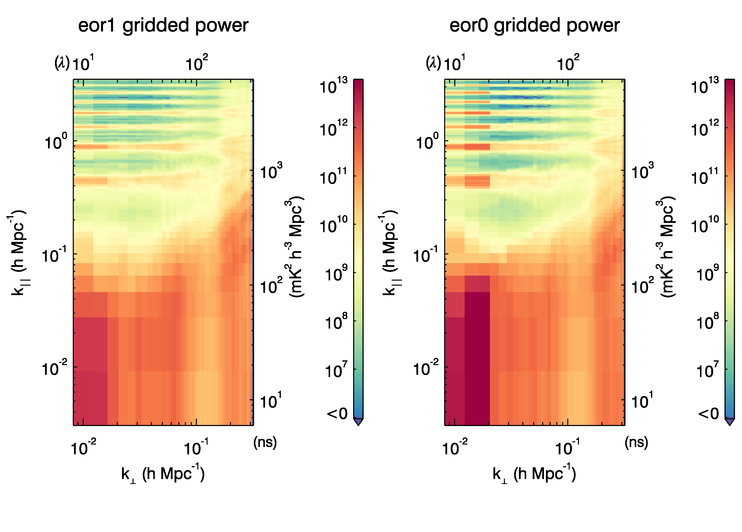}
\caption{Gridded power spectra for the 21 hours of observations on two fields used in this work, as processed through the CHIPS estimator \citep{trottchips2016}.}
 \label{fig:chips_ps}
\end{center}
\end{figure*}

\begin{figure}
\begin{center}
\includegraphics[width=18pc]{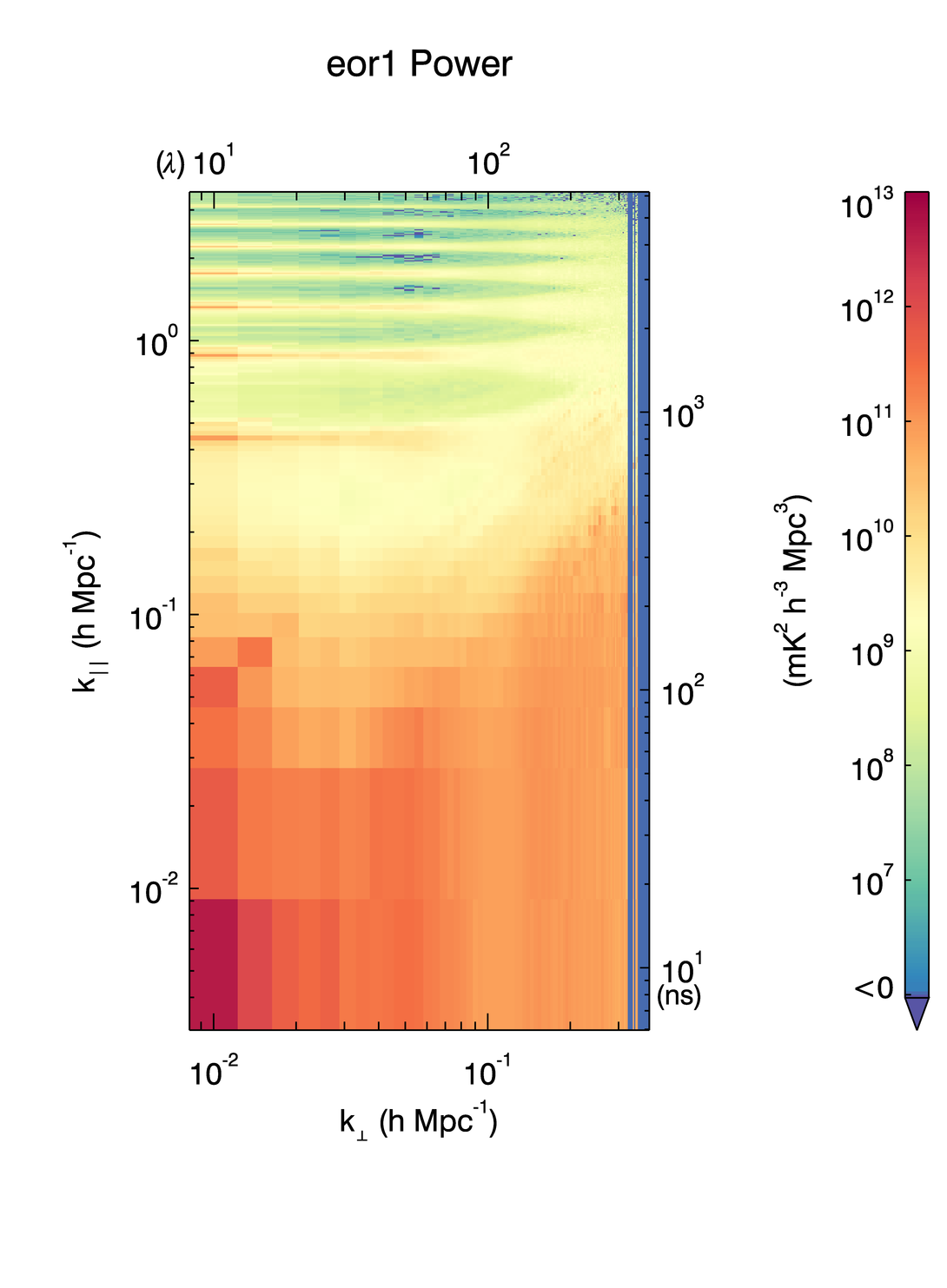}
\caption{Delay transform power spectra for the EoR1 field for the data used in this analysis. Note the large leakage into the EoR window, which yields large denominators for the normalised bispectrum.}
 \label{fig:eor0_ps}
\end{center}
\end{figure}

\begin{figure}
\begin{center}
\includegraphics[width=18pc]{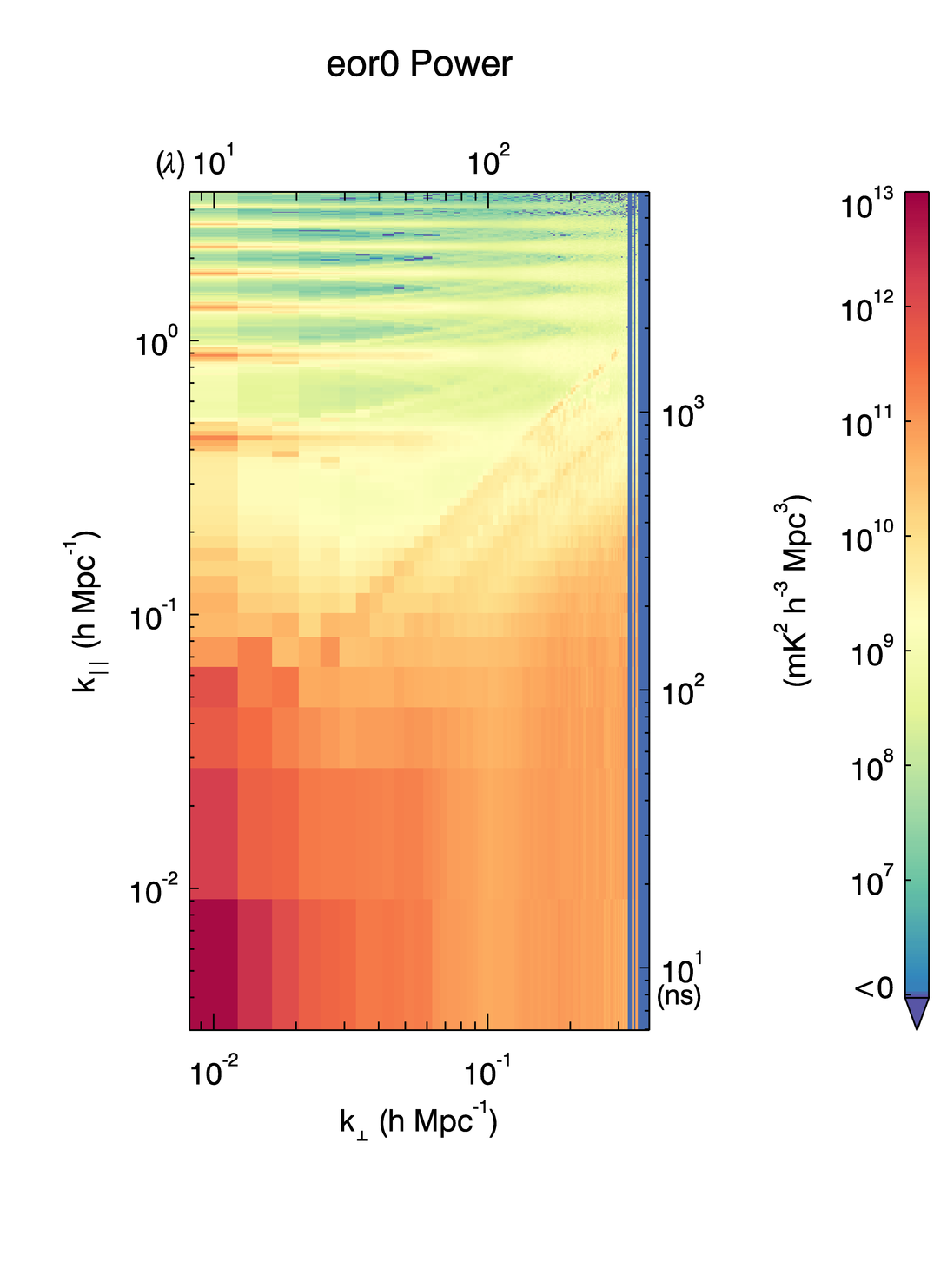}
\caption{Delay transform power spectra for the EoR0 field for the data used in this analysis.}
 \label{fig:eor1_ps}
\end{center}
\end{figure}

The normalised bispectrum, $\mathcal{B}$, is normalised by the power spectra at each of the $k$ modes forming the triangles. Figure \ref{fig:chips_ps} shows the power spectra for the EoR1 and EoR0 fields for the full datasets as used in the gridded estimator. These have been processed through the CHIPS power spectrum estimator \citep{trottchips2016}.
Figures \ref{fig:eor0_ps} and \ref{fig:eor1_ps} show the corresponding delay spectra, as used in the direct bispectrum.
There are small differences between the two power spectrum estimators, as is expected given that delay spectra do not grid with primary beams, and Fourier Transform along frequency, yielding different results for longer baselines. The signature of Galactic emission from close to the horizon is evident in the EoR0 power spectra, while it is less structured in EoR1, where the Galactic Centre has set. Most notably, the delay spectra show large foreground leakage into the EoR window ($k_\parallel<0.4$), yielding large power spectrum denominator values for the normalised bispectrum.

Using these data, Table \ref{table:reduced_bispec} describes the normalised bispectrum.
\begin{table*}
\centering
\begin{tabular}{|c||c|c||c|}
\hline 
Triangles & Direct & Gridded & Type \\
\hline
{\bf EoR0} & & & {\bf 14m}\\
\hline
$k_1=k_2=k_3=0.007$ & $0.166 \pm 2.5e-7$ & $10.4 \pm 4.1e-8$ & Equilateral \\
$k_1=0.2,k_2=k_3=0.1$ & $-0.266 \pm 0.0004$ & $921.2 \pm 0.3$ & Isosceles \\
$k_1=0.4,k_2=k_3=0.2$ & $2.84 \pm 0.0044$ & $-1766.2 \pm 0.6$ & Isosceles \\
$k_1=0.6,k_2=k_3=0.3$ & $-4.87 \pm 0.063$ & $-427.8 \pm 5.8$ & Isosceles \\
$k_1=1.0,k_2=k_3=0.5$ & $3.45 \pm 0.60$ & {$\bf 129.4 \pm 108.9$} & Isosceles \\
\hline
{\bf EoR0} & & & {\bf 28m}\\
\hline
$k_1=k_2=k_3=0.014$ & $-0.019 \pm 1.4e-7$ & $-29.3 \pm 8.1e-6$ & Equilateral \\
$k_1=0.2,k_2=k_3=0.1$ & $-0.14 \pm 0.002$ & $-594.5 \pm 4.0$ & Isosceles \\
$k_1=0.4,k_2=k_3=0.2$ & $0.360 \pm 0.009$ & $948.9 \pm 7.2$ & Isosceles \\
$k_1=0.6,k_2=k_3=0.3$ & {$\bf 0.98 \pm 0.18$} & $-793.1 \pm 37.6$ & Isosceles \\
$k_1=1.0,k_2=k_3=0.5$ & {$\bf 1.08 \pm 1.78$} & $19450 \pm 752$ & Isosceles \\
\hline \hline
{\bf EoR1} & & &  {\bf 14m} \\
\hline
$k_1=k_2=k_3=0.007$ & $-0.004 \pm 1.2e-8$ & $0.61 \pm 3.2e-9$ & Equilateral \\
$k_1=0.2,k_2=k_3=0.1$ & $0.044 \pm 0.0001$ & $-666.5 \pm 0.03$ & Isosceles \\
$k_1=0.4,k_2=k_3=0.2$ & $0.19 \pm 0.0004$ & $3157.0 \pm 0.82$ & Isosceles \\
$k_1=0.6,k_2=k_3=0.3$ & $-0.064 \pm 0.007$ & $-1861.9 \pm 0.54$ & Isosceles \\
$k_1=1.0,k_2=k_3=0.5$ & {$\bf -0.12 \pm 0.13$} & $5907.5 \pm 56.1$ & Isosceles \\
\hline
{\bf EoR1} & & &  {\bf 28m} \\
\hline
$k_1=k_2=k_3=0.014$ & $0.0006 \pm 7.0e-9$ & $17.1 \pm 8.4e-7$ & Equilateral \\
$k_1=0.2,k_2=k_3=0.1$ & {$\bf 0.0001 \pm 0.0005$} & $927.7 \pm 0.43$ & Isosceles \\
$k_1=0.4,k_2=k_3=0.2$ & $-0.082 \pm 0.002$ & $-245.3 \pm 0.15$ & Isosceles \\
$k_1=0.6,k_2=k_3=0.3$ & {$\bf 0.012 \pm 0.030$} & $5881.8 \pm 10.4$ & Isosceles \\
$k_1=1.0,k_2=k_3=0.5$ & {$\bf 6.2 \pm 1.2$} & $4257.6 \pm 15.6$ & Isosceles \\
\hline \hline
\end{tabular}
\caption{Normalised bispectrum estimates, $\mathcal{B}$, and one sigma uncertainties for the direct and gridded bispectra for each observing field and triangle type. Bold-faced values indicate bispectrum estimates that are consistent with thermal noise.}\label{table:reduced_bispec}
\end{table*}
Bold-faced results are broadly consistent with thermal noise ($<5\sigma$), again reflecting the modes that are least affected by foregrounds. The difference between the dimensional and reduced bispectrum results is due to the different power spectral estimators. Also notable is the difference in amplitude of the gridded and direct normalised bispectrum estimates. Due to the division by the power spectrum, the normalised bispectrum is heavily-dependent on the details of the power spectrum estimates, which fluctuate substantially in foreground-affected regions. The delay-space power spectra show increased foreground power in the EoR window, and this is reflected in a larger power spectrum estimate, and therefore a lower normalised bispectrum. This reliance highlights the complexity for interpreting the normalised bispectrum with foreground-affected data.

\section{Bispectrum signature of foregrounds}\label{sec:fg}
Estimates of bispectrum sensitivity for operational and future 21~cm experiments are incomplete without a treatment of foregrounds. Despite the expectation that point source, continuum foregrounds only impact a region of the three dimensional EoR parameter space ($k_x,k_y,k_\parallel$), in reality the details of the instruments, complexity of extragalactic and Galactic emission, limited bandwidth and calibration errors leave residual contaminating signal throughout the full parameter space. Although these methods perform very well to remove such signal, the extreme dynamic range demanded by this experiment translate to bias that exceeds the expected cosmological signal strength. The results presented here are clearly foreground-dominated, particularly for the equilateral triangle configuration.

As such, the bispectrum signature of foregrounds can be computed for a simple point source foreground model. {We first consider the expected foreground bispectrum, which quantifies the bias in the measurement, and then turn to the variance of the foreground bispectrum, which quantifies the additional noise term}.

We employ a model where the sky is populated with a random distribution of unresolved extragalactic point sources that follow a low-frequency number counts distribution \citep{intema11,franzen16}:
\begin{equation}
    \frac{dN}{dS} = \alpha S^\beta \,\, {\rm Jy}^{-1} \rm{sr}^{-1},
\end{equation}
where $\alpha \simeq 3900$ and $\beta=-1.59$ for sources with flux density at 150~MHz of less than 1 Jansky. We assume there is no source clustering and spectral dependence, yielding a Poisson-distributed number of sources in each differential sky area.

{The clustering of point sources in the power spectrum has been studied by \citet{murray17}. They find that source clustering will be unimportant for the MWA (unless the clustering is extreme, which is not measured), but may be important for the SKA, which can clean to deeper source levels. Nonetheless, the structure due to clustered point source foregrounds only changes the amplitude of the foreground structure in the EoR wedge as a function of angular scale ($k_\bot$). Because the line-of-sight spectral component is unaffected, the signature of clustered foregrounds in the EoR Window is mostly unchanged. These more realistic point source foregrounds will be considered in the simulations of \citet{cathcath18} and here we retain the analytic signature of the Poisson foregrounds.}

We further assume that the primary beam can be approximated by a frequency-dependent Gaussian:
\begin{equation}
    A(l,m,\nu_0) = \exp{\left(-\frac{(l^2+m^2)\nu_0^2{\rm A_{eff}}}{2c^2\epsilon^2} \right)},
    \label{eqn:beam}
\end{equation}
where A$_{\rm eff}$ is the tile effective area, and $\epsilon$ encodes the conversion from an Airy disk to a Gaussian.

The visibility is given by Equation \ref{eqn:vis} for frequency $\nu$. To compute the line-of-sight component to the visibility, we Fourier Transform over frequency channels, after employing a frequency taper (window function) to reduce spectral leakage from the finite bandwidth:
\small
\begin{eqnarray}
    V(u,v,\eta) &=& \int dldm S(l,m,\nu_0) A(l,m,\nu_0) \nonumber \\
    &\times& \int d\nu \Upsilon(\nu) \exp{\left(-2\pi{i}\frac{\nu(xl+ym)}{c}\right)} \exp{\left(-2\pi{i}\nu\eta\right)} \nonumber \\
    &=& \int dldm S(l,m,\nu_0) A(l,m,\nu_0) \nonumber \\
    &\times& \int d\nu \Upsilon(\nu) \exp{\left(-2\pi{i}\nu(xl/c+ym/c+\eta)\right)}\\
    &=& \int dldm S(l,m) A(l,m) \nonumber\\
    &\times& \tilde{\Upsilon}(xl/c+ym/c+\eta) \,\, {\rm Jy Hz},
\end{eqnarray}
\normalsize
whee $\Upsilon(\nu)$ is the spectral taper, and we have performed the Fourier Transform over frequency. For analytic tractability, in this work we use a Gaussian taper, with a characteristic width, $\Sigma \simeq $BW/7, such that the edges of the band are consistent with zero and it is well-matched to a Blackman-Harris taper:
\begin{equation}
    \Upsilon(\nu) = \exp{-\frac{\nu^2}{2\Sigma^2}},
\end{equation}
with corresponding Fourier Transform,
\begin{equation}
    \tilde{\Upsilon}(\eta) = \sqrt{2\pi\Sigma^2} \exp{-2\pi^2\Sigma^2\eta^2} {\rm Hz}.
\end{equation}

The bispectrum is formed from the triple product of visibilities. Accounting for the fact that the point sources are only correlated locally ($\delta_D(l_1+l_2+l_3=0)$), its expected value with respect to foregrounds is:
\small
\begin{equation}
    \langle V_1V_2V_3\rangle = \int dldm \langle S^3(l,m) \rangle A^3(l,m) \exp{\left(-2\pi^2\Sigma^2T^2\right)}
    \label{eqn:triplet}
\end{equation}
where,    
\begin{eqnarray}
    T^2 &=& \left(\frac{x_1l}{c}+\frac{y_1m}{c}-\eta_1 \right)^2 \nonumber \\
    &+& \left(\frac{x_2l}{c}+\frac{y_2m}{c}-\eta_2 \right)^2 + \left(\frac{x_3l}{c}+\frac{y_3m}{c}-\eta_3 \right)^2.
\end{eqnarray}
\normalsize
Here, the source counts have been separated from the spatial integral. This is a general expression for a triplet of baselines. We can now simplify this for triangles, particularly those with isosceles configurations (where the equilateral is a single case of an isosceles).

Closed triangles follow the relations:
\begin{align}
    x_1 + x_2 = -x_3\\
    y_1 + y_2 = -y_3\\
    \eta_1 + \eta_2 = -\eta_3,
\end{align}
and we define, without loss of generality, the following relations for the isosceles configurations considered in this work:
\begin{eqnarray}
    x_1 &=& -2x_2 \nonumber\\
    x_2 &=& x_3 \nonumber\\
    y_1 &=& 0 \nonumber\\
    y_2 &=& -y_3 \nonumber\\
    y_2 &=& x_1\cos{\pi/6} = 2x_2\cos{\pi/6} = \sqrt{3}x_2 \nonumber\\
    2\eta_2 &=& 2\eta_3 = -\eta_1.
\end{eqnarray}
Making these substitutions in Equation \ref{eqn:triplet}, completing the squares and collecting terms, we find:
\begin{eqnarray}
    \langle V_1V_2V_3 \rangle = \int dldm \langle S^3(l,m) \rangle A^3(l) \nonumber\\
    \times \exp{\left(-12\pi^2\Sigma^2 \left(x_2^2/c^2(l^2+m^2) + \eta_2^2 \right)\right)}.
    \label{eqn:triplet_iso}
\end{eqnarray}
The source count expectation value uses the source number counts distribution and the fact that the number of sources at any sky location is Poisson-distributed to find:
\begin{eqnarray}
    \langle S^3(l,m) \rangle &=& \int_S S^3(\nu_0) \frac{dN}{dS}\nonumber\\
    &=& \frac{\alpha}{4+\beta} S_{\rm max}^{4+\beta} \,\, {\rm Jy^3 sr^{-1}}
\end{eqnarray}

Incorporating the primary beam from Equation \ref{eqn:beam}, moving to polar coordinates, and performing the integral over $(l,m)$, we find {for the expected foreground bispectrum bias}:
\begin{eqnarray}
    \langle V_1V_2V_3 \rangle = (2\pi\Sigma^2)^{1.5} \frac{\alpha}{4+\beta} S_{\rm max}^{4+\beta}  \nonumber\\
    \times \frac{\pi}{\theta}\exp{\left(-\frac{\pi^2\rm{BW}^2\eta_2^2}{25}\right)} \,\, {\rm Jy^3 Hz^3}
    \label{eqn:triplet_iso}
\end{eqnarray}
where,
\begin{equation}
    \theta = \frac{3{\rm A_{eff}\nu_0^2}}{c^2} + \frac{\pi^2{\rm BW}^2u_2^2}{25\nu_0^2},
\end{equation}
and BW is the experiment bandwidth. This factor combines the primary beam (spatial taper) and spectral taper components into a single factor.

The equilateral configuration can be derived from this expression with $\eta_2=0$. For the 28~m baselines, a maximum source flux density of 1~Jy and A$_{\rm eff}=21$~m$^2$, and performing the cosmological conversions, we expect a bispectrum estimate of:
\begin{equation}
    B(x=28) \simeq 8.6 \times 10^{19} {\rm mK}^3 h^{-6} {\rm Mpc}^6,
\end{equation}
which is comparable to the estimates found in Section \ref{sec:results}. For 14~m baselines, $B(x=14) \simeq 1.0 \times 10^{20} {\rm mK}^3 h^{-6} {\rm Mpc}^6$.

\begin{figure}
\begin{center}
\includegraphics[width=20pc]{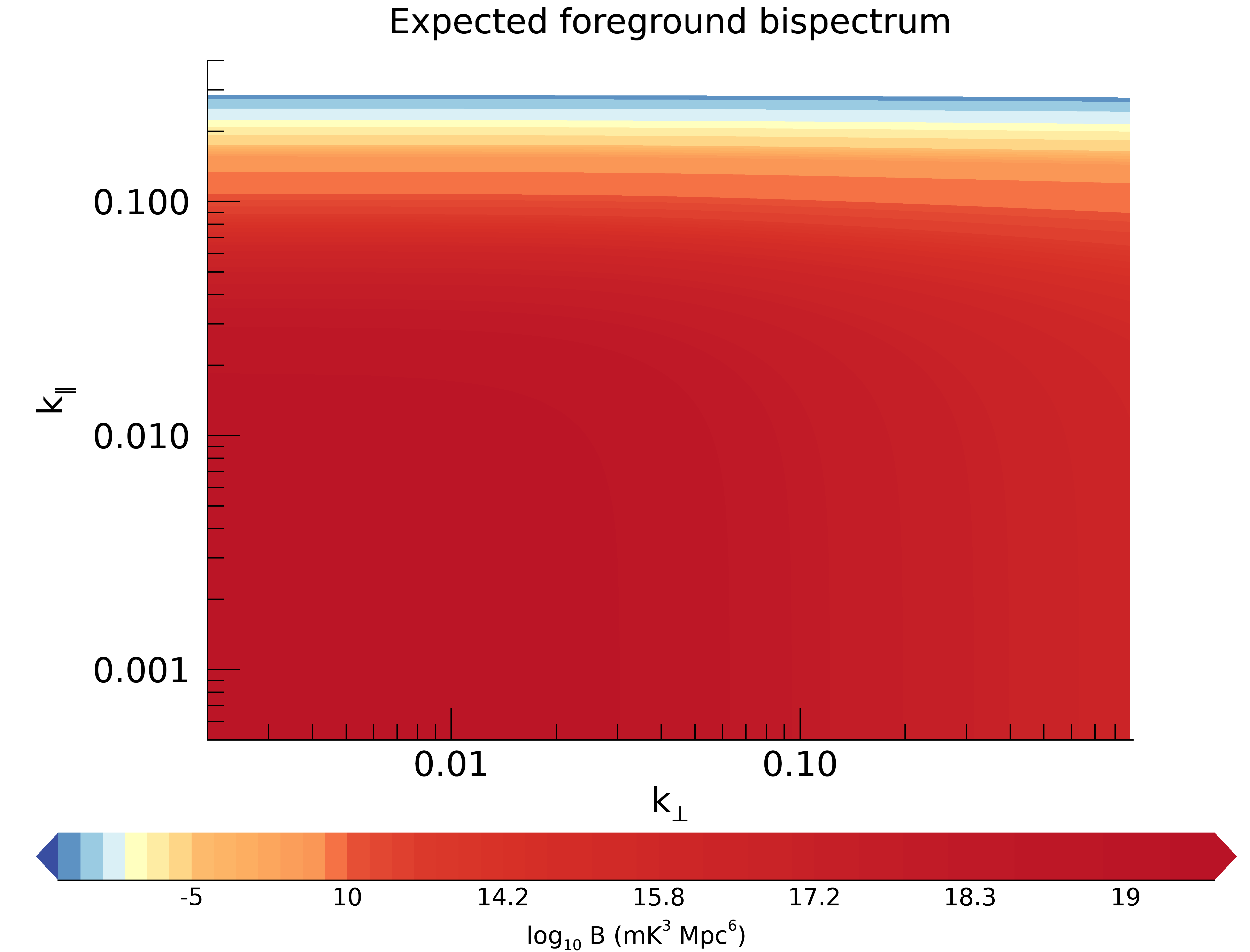}
\caption{Point source-foreground dimensional bispectrum signature of isosceles triangle vectors in $k_\bot-k_\parallel$-space (note the stretched logarithmic colour bar). In this model, the expected foreground signal has fallen to below the expected cosmological signal value by $k_\parallel \geq 0.12$.}
 \label{fig:fg}
\end{center}
\end{figure}

The isosceles configurations incorporate the $\eta$ term. For $k_\parallel>0.1$, this term decays to below the noise, which is consistent with that observed in the data. The signature of this isosceles foreground dimensional bispectrum in $k_\bot$-$k_\parallel$ space is shown in Figure \ref{fig:fg}.
For the $k_1=0.1 h$~Mpc$^{-1}$ stretched configuration, we expect for 28~m (14~m):
\begin{equation}
    B \simeq 1.7 \times 10^{12} \,\, (1.0 \times 10^{12}) {\rm mK}^3 h^{-6} {\rm Mpc}^6.
\end{equation}
The squeezed configurations of large $k_\bot$ combined with small $k_\parallel$ might be interesting for future studies, depending on the expected cosmological signal on these scales. Given that the power spectrum is expected to be small on these {combination of line-of-sight and angular} scales, most EoR experiments are not designed for high sensitivity here ($k_\bot=0.1$ corresponds to 200~m baselines).

Interestingly, the expected foreground bispectrum signal is positive, due to its constituent astrophysical sources being associated with overdensities. Conversely, the stretched isosceles 21-cm bispectrum from the cosmological signal will be negative on many scales during reionisation \citep{majumdar18}.

\subsection{Normalised foreground bispectrum}
The normalised bispectrum also contains the expected power spectrum values for a foreground model. In line with the methodology developed in the previous section, we can write the expected power spectrum at $(u,v,\eta)$ as:
\begin{eqnarray}
        P(u,v,\eta) &=& \langle V^\ast(u,v,\eta) V(u,v,\eta) \rangle \nonumber\\
        &=& (2\pi\Sigma^2) \frac{\alpha}{3+\beta} S_{\rm max}^{3+\beta} \nonumber\\ &\times& \left( \rm{erf}\left(\frac{b+2a}{\sqrt{2}a}\right) - \rm{erf}\left(\frac{b-2a}{\sqrt{2}a} \right) \right) \nonumber\\
        &\times& \exp{-4\pi^2\Sigma^2\eta^2} \sqrt{\frac{\pi}{4a}} \exp{\frac{b^2}{4a}},
\end{eqnarray}
where
\begin{eqnarray}
        a &=& \frac{2\pi c^2}{\nu_0^2{\rm A_{eff}}/\epsilon^2} + \frac{4\Sigma^2 |x|^2}{c^2}\\
        b &=& \frac{8\Sigma^2|x|\eta}{c},
\end{eqnarray}
encode the spatial and spectral tapers, and $|x|^2=x^2+y^2$ (without loss of generality). This expression is derived from the Fourier Transform over Gaussians, and then the integral over $dldm$\footnote{This can also be derived as a covariance between $u$ modes and $\eta$ modes, which encodes the spectral leakage that stems from the spatial and spectral tapers. This covariance is that used to understand power spectrum uncertainties in EoR work, where correlations between $k$-cells must be correctly treated.}. 

When $\eta=0$ and for the 28~m baseline triangles, the expected bispectrum normalisation is:
\begin{equation}
    \sqrt{P(u,v,\eta)^3}/V = 2.8 \times 10^{21} \,\, {\rm mK^3 h^{-6}  Mpc^6}.
\end{equation}
For the 14~m triangles, we find, $\sqrt{P(u,v,\eta)^3}/V = 3.3 \times 10^{21} \,\, {\rm mK^3 h^{-6}  Mpc^6}.$ 
When compared with the expected bispectrum value, we find that (28~m):
\begin{equation}
    \langle \mathcal{B} \rangle = 1.7,
\end{equation}
and $\langle \mathcal{B} \rangle = 0.6$ for the 14~m baselines, which exceed the equilateral triangle configuration estimates from the MWA data. As with the bispectrum estimate, the isosceles configurations have expected power values that fall rapidly with $\eta$, and are less comparable to the data in these idealised scenarios. However, for the $k_1=0.1 h$~Mpc$^{-1}$ stretched configuration, we expect for 28~m (14~m):
\begin{equation}
    \langle \mathcal{B} \rangle = 4.0 \,\, (240,000).
\end{equation}
These values are ratios of very small numbers, and therefore are highly dependent on numerical details and are not representative. However, they may lend support to the idea that the normalised bispectrum is difficult to interpret because it relies on foreground details in both the bispectrum and power spectrum. 

Alternatively, the combination of power spectrum, dimensional bispectrum and normalised bispectrum may help to shed additional light on whether data are really foreground free. Given the different behaviour of foregrounds in these statistics, this information may be used to discriminate cosmological information from foregrounds, or to help to design some iterative foreground cleaning algorithm, taking into consideration their behaviour in cosmological simulations. In this scenario, the normalised bispectrum may provide useful information.

\begin{figure}
\begin{center}
\includegraphics[width=20pc]{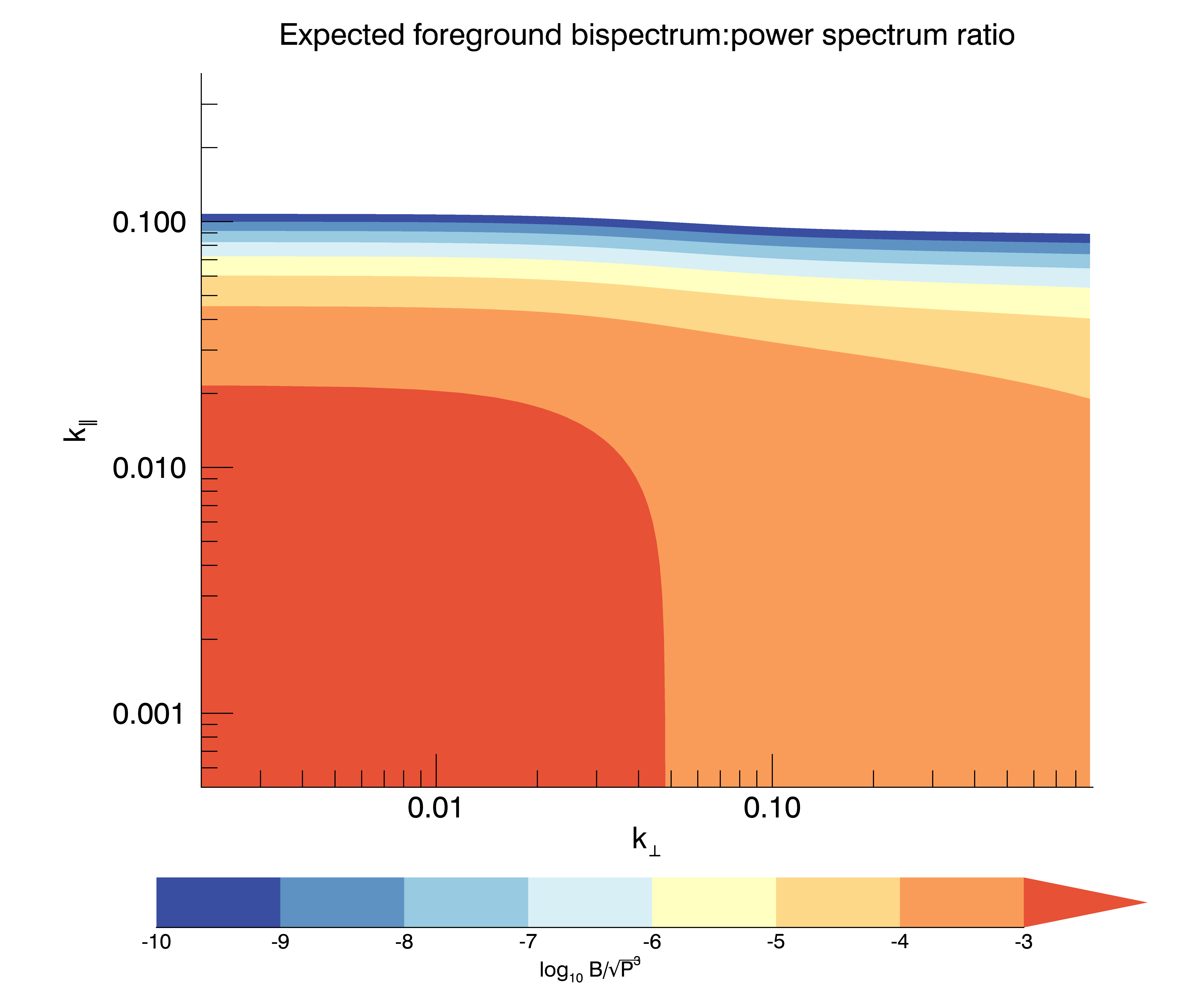}
\caption{Point source- normalised foreground bispectrum signature of isosceles triangle vectors in $k_\bot-k_\parallel$-space (note the stretched logarithmic colour bar). In this model, the expected foreground signal has fallen to below the expected cosmological signal value by $k_\parallel \geq 0.12$.}
 \label{fig:bispec_powerspec_ratio}
\end{center}
\end{figure}

{Figure \ref{fig:bispec_powerspec_ratio} displays the normalised foreground bispectrum for isosceles configurations in $k_\bot-k_\parallel$ space.}
{For the point source foregrounds, the power spectrum denominator dominates over the expected bispectrum signal, yielding values $<10^{-3}$ across all of parameter space. This presents an interesting divergence from the usual expectation of foreground bias in the power spectrum, where foregrounds add overall signal. In this case, a large measurement that exceeds the thermal noise is consistent with a cosmological origin, and not with residual foregrounds.}

\subsection{Foreground bispectrum error}\label{sec:fg_var}
{We now turn our attention to consideration of the signal variance due to residual foregrounds, $\langle B_{\rm FG}^2\rangle$, such that (cf Equation \ref{eqn:errorbispec})}:
\begin{equation}
    \Delta{B}^2 = \frac{{3}\sigma_{\rm therm}^6}{{\displaystyle\sum_j W_j}} + \langle B_{\rm FG}^2\rangle,
\end{equation}
and
\begin{equation}
    \langle B_{\rm FG}^2\rangle = \langle  V_1^{\ast}V_2^{\ast}V_3^{\ast}V_1V_2V_3  \rangle.
\end{equation}
{This reduces to a relatively simple expression for the simple point source case, due to the cancelling of complex components (this is not generally true for the covariance). Using the same formalism as earlier, and again considering the Poisson-distributed nature of the flux density of the sources, we find:}
\begin{eqnarray}
    {\langle B_{\rm FG}^2\rangle} &=&  \int S^6\frac{dN}{dS} dS \int A^6(l,m) dldm \\
    &\times& \int \vec\Upsilon {\rm e}^{(-2\pi{i}(\eta_1\Delta\nu_{12}+\eta_2\Delta\nu_{34}+\eta_3\Delta\nu_{56})} d\vec{\nu}\nonumber\\
    &=& \alpha\frac{S_{\rm max}^{7+\beta}}{7+\beta} \left(12\pi \frac{c^2\epsilon^2}{\nu_0^2A_{\rm eff}}\right) {\rm e}^{-4\pi^2\Sigma^2(\eta_1^2+\eta_2^2+\eta_3^2)},\nonumber\\
    &=& \alpha\frac{S_{\rm max}^{7+\beta}}{7+\beta} \left(12\pi \frac{c^2\epsilon^2}{\nu_0^2A_{\rm eff}}\right) {\rm e}^{-6\pi^2\Sigma^2\eta_1^2},\nonumber
\end{eqnarray}
{where $\vec\Upsilon \equiv \Upsilon(\nu_1)\Upsilon(\nu_2)\Upsilon(\nu_3)\Upsilon(\nu_4)\Upsilon(\nu_5)\Upsilon(\nu_6)$. This expression is flat in angular modes, and decays rapidly in line-of-sight modes.}

\begin{figure}
\begin{center}
\includegraphics[width=24pc]{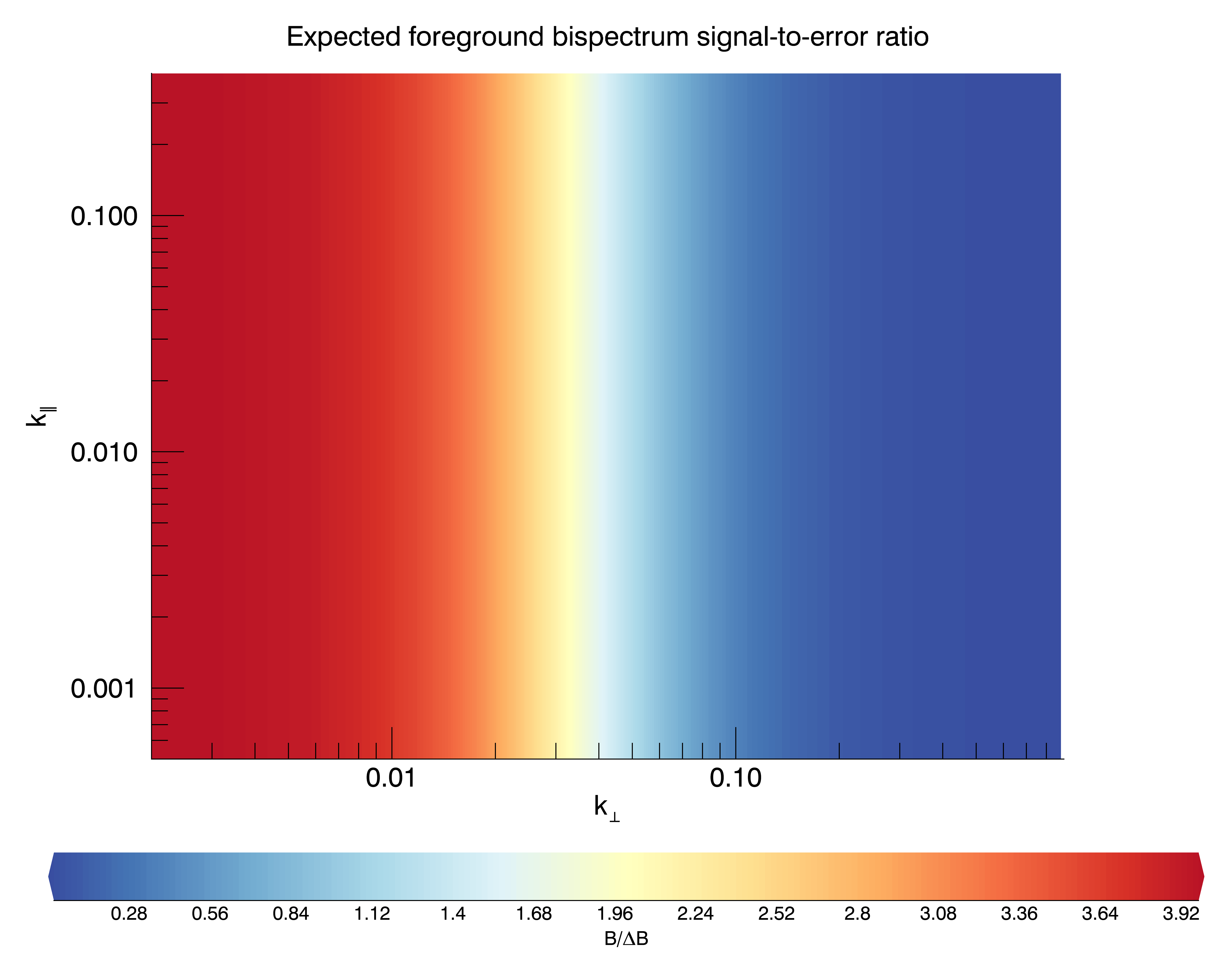}
\caption{Ratio of point source foreground bispectrum bias to uncertainty, for isosceles triangle vectors in $k_\bot-k_\parallel$-space (note linear plot). The bias exceeds the uncertainty at large angular modes, but rapidly falls below for larger $k_\bot$, with no dependence on line-of-sight scale.}
 \label{fig:snr_fg}
\end{center}
\end{figure}

{Comparing this with the expected value of the foreground bispectrum, Equation \ref{eqn:triplet_iso}, we can form the foreground bispectrum signal-to-error ratio; Figure \ref{fig:snr_fg}.}
{The signal bias exceeds the uncertainty for small scales, but on the larger scales of interest for EoR, the uncertainty dominates. Nonetheless, there is no line-of-sight dependence, demonstrating that the foreground bias and uncertainty both drop rapidly and are negligible for $k_\parallel>0.12 h$Mpc$^{-1}$, implying that for larger $k_\parallel$ scales, point source foregrounds are not significant in the signal or noise budget.}

\begin{figure}
\begin{center}
\includegraphics[width=20pc]{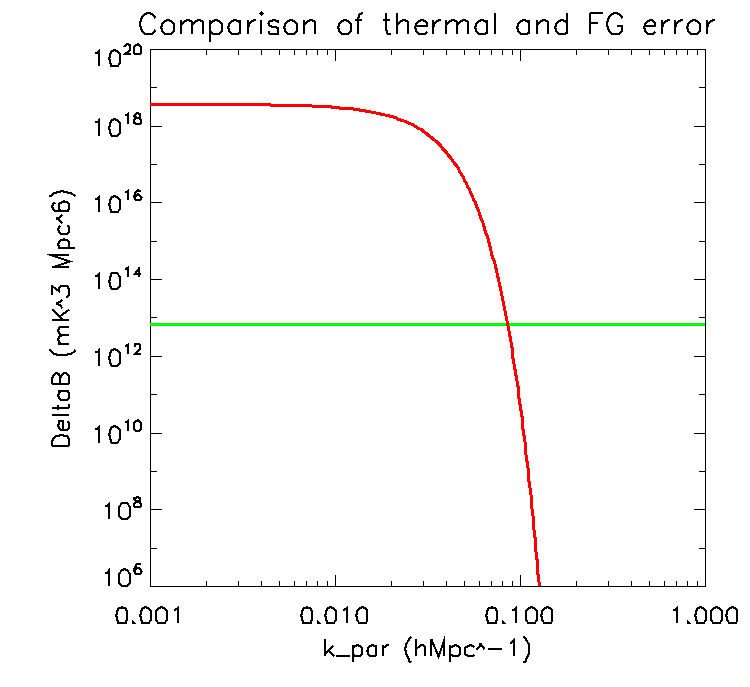}
\caption{Errors for point source-normalised foreground bispectrum (red) and thermal noise (green), for isosceles triangle vectors in $k_\bot-k_\parallel$-space and 300 observations used in this work for the EoR1 field.}
 \label{fig:fg_thermal}
\end{center}
\end{figure}

{One can also now compare the foreground uncertainty to the expected thermal noise level. For the EoR1 field data, the measured uncertainty for the direct bispectrum estimator was 6.7 $\times$ 10$^{12}$ mK${^3}$ Mpc$^6$. Figure \ref{fig:fg_thermal} shows this level (green line) compared with the foreground bispectrum error (red line) as a function of line-of-sight scale. (The gridded estimator has slightly lower noise level, but the distinction is not significant when compared to the large gradient of the foreground contribution.) As with the foreground bias, the error induced by residual foregrounds drops steeply beyond $k_\parallel=0.12 h$Mpc$^{-1}$, and falls below the thermal noise (even in this case with a small dataset for the EoR1 field).}

{As a final step to assessing the advantages of the bispectrum compared with the power spectrum to detect the cosmological signal, we divide the expected 21~cm bispectrum values for the faint and bright galaxies, presented in Table \ref{table:bispectra} by the foreground error for these modes, and compare with that for the power spectrum. The power spectrum values for faint and bright galaxies are taken from the same underlying dataset generated by 21cmFAST \citep{mesinger11}. For all but the $k_1=0.1 h$Mpc$^{-1}$ mode, the foreground uncertainty is negligible, and the ratio is uninteresting. For $k_1=0.1 h$Mpc$^{-1}$, $k_2=k_3=0.2 h$Mpc$^{-1}$:}
\begin{eqnarray}
    P_{21} = 1.3 \times 10^4 mK^2 Mpc^3\nonumber\\
    \Delta{P}_{FG} = 0.2 \times 10^{-1} mK^2 Mpc^3\nonumber\\
    B_{21} = 4.4 \times 10^9 mK^3 Mpc^6\nonumber\\
    \Delta{B}_{FG} = 4.6 \times 10^{10} mK^3 Mpc^6,
\end{eqnarray}
{yielding better performance for the power spectrum, within the foreground dominated region.}

{However, outside of the foreground `wedge', which exists in both power spectrum and bispectrum space, the data uncertainty is limited by the thermal noise, and here the bispectrum achieves higher signal-to-noise ratio for a set observation time:}
\begin{eqnarray}
    P_{21} = 1.3 \times 10^4 mK^2 Mpc^3\nonumber\\
    \Delta{P}_{\rm therm} = 5.7 \times 10^6 mK^2 Mpc^3\nonumber\\
    B_{21} = 4.4 \times 10^9 mK^3 Mpc^6\nonumber\\
    \Delta{B}_{\rm therm} = 5.3 \times 10^{11} mK^3 Mpc^6.
\end{eqnarray}
{Taking the ratios we find,}
\begin{eqnarray}
    P_{21}/\Delta{P}_{\rm therm} = 0.002\nonumber\\
    B_{21}/\Delta{B}_{\rm therm} = 0.008\nonumber.
\end{eqnarray}
{\textit{Accounting for the fact that the gridded bispectrum averages down with $t^{1.5}$ while the gridded power spectrum averages with $t$}, the observing time multiple (above 10 hours) for a detection (SNR=1) is}
\begin{eqnarray}
    t_{P} = 500 \times\nonumber\\
    t_B = 25 \times\nonumber.
\end{eqnarray}
{Therefore, the bispectrum detection can theoretically be achieved in a fraction of the time of the power spectrum detection, for thermal noise-limited modes close to the EoR wedge. A SNR=1 detection level could potentially be reached in 250 hours, \textit{for this wave mode.} This conclusion is relevant for the MWA, where the excellent instantaneous $uv$-coverage allows for rapid observation of triangle configurations. \citep[We note that the power spectrum SNR shown here is not inconsistent with previous expectations for the performance of the MWA, because it applies only to this single mode;][]{beardsley16,wayth18}. Future work presented in \citet{cathcath18} will explore a more full range of triangle configurations and foreground bias and error.}


\section{Discussion and conclusions}
As discussed, the model used for the foreground bispectrum signal predicts that isosceles configurations have amplitudes that fall rapidly with non-zero $k_\parallel$. Nevertheless, we find that the numerator and denominator of the normalised bispectrum scale such that its amplitude in this model increases as a function of $\eta$. For the 28~m baselines, the ratio doubles by $k_\parallel > 0.014 h$~Mpc$^{-1}$. Over the same $k$ range, the dimensional bispectrum is rapidly decaying, losing seven orders of magnitude from the $k_\parallel=0$ mode. In this model, the expected foreground signal has fallen to below the expected cosmological signal value by $k_\parallel \geq 0.12$. The results from the MWA datasets have some modes thermal noise-limited at 10~hours, and only the $k_\parallel=0$ mode is clearly foreground dominated for all experiments. In line with the discussion of \citet{bharadwaj05}, it is possible that the dimensional bispectrum is less affected by foregrounds than the power spectrum. However, the normalised bispectrum is more difficult to interpret, given the different observational foreground effects on the bispectrum numerator and power spectra denominator.

Reduction of foreground contamination is an active field of research in 21~cm EoR experiments, and primary motivator for testing statistics other than the power spectrum. Despite the normalised bispectrum providing a cosmologically stable and robust estimate of non-Gaussianity compared with the dimensional bispectrum, the expected foreground value is difficult to discriminate from the expected signal value \citep{watkinson18}. Conversely, the dimensional bispectrum yields values that are highly-significant detections, showing clear foreground contamination. Thus, the normalised bispectrum may not be the best discriminant in real EoR experiments. For future experiments, with higher sensitivity, exploration of modes with negative bispectra may help discrimination from foreground contamination, where the bispectrum is expected to be positive This is explored further in \citet{cathcath18} and demonstrated previously in \citet{lewis11}. It would also be interesting to study the signature of calibration errors in radio data on bispectrum estimates, to explore whether they have an imprint that can be discriminated from the cosmological signal.

The thermal noise levels, as are achieved in these 10 hour datasets for large $k$ isosceles configurations, are 3--4 orders of magnitude larger than the expected bispectrum value for these configurations at low redshifts \citep{majumdar18}. The gridded bispectrum noise scales with observation time to the power of 1.5, requiring a 1000~h observation with the MWA to achieve a cosmological detection. This estimate is in line with predictions from \citet{yoshiura15}. Further advantage may be gained from incoherent addition of isosceles triangle configurations with similar vector lengths, where the bispectrum is expected to vary slowly with changing parameters. An initial test of this for the $k_1=0.1 h$~Mpc$^{-1}$ mode shows an improvement in sensitivity by a factor of ten for the gridded estimator, yielding a theoretical detection of the signal with 150~hours of data. The direct estimator scales incoherently with time ($t^{0.75}$), due to the incoherent addition of triangles from different observations, but does utilise coherent addition of instantaneously-redundant triads.

We have presented the first effort to estimate the cosmological bispectrum from the Epoch of Reionisation with 21~hours of MWA data, and have shown the parts of parameter space that are consistent with thermal noise at this level, using two types of bispectrum estimator. These two approaches are presented in order to demonstrate the practicalities of estimation of the bispectrum with real radio interferometer data. We have also derived a form for the expected bispectrum signature of point source foregrounds for equilateral and isosceles configurations, and demonstrated broad consistency between the analytic model and the estimates obtained from the data. 

By considering the foreground bispectrum variance in the noise estimation, we have demonstrated that both the foreground bias and variance are insignificant for $k_\parallel<0.12h$~Mpc$^{-1}$, allowing these regions of parameter space to be probed with dominant thermal noise. Due to the ability of the gridded bispectrum estimator to reduce thermal noise proportional to $t^{1.5}$, unlike the power spectrum which reduces with $t$, the 21~cm cosmological bispectrum may be detectable with fewer observing hours than the power spectrum for arrays with excellent instantaneous $uv$-coverage (i.e., with well-sampled baselines). This insight makes observational pursuit of the bispectrum worthwhile for some current instruments.

In a companion paper \citep{cathcath18}, we explore optimal triangles to study from a signal and foreground contamination ratio perspective. Future work can also study the signature of calibration errors on the bispectrum. This work helps to define the optimal observational strategy and approach to bispectrum studies.

\begin{acknowledgements}
This research was supported by the Australian Research Council Centre of Excellence for All Sky Astrophysics in 3 Dimensions (ASTRO 3D), through project number CE170100013. CMT is supported by an ARC Future Fellowship under grant FT180100196.
CAW and SM acknowledge financial support from the European Research Council under ERC grant number 638743-FIRSTDAWN (held by Jonathan R. Pritchard).
KT's work is partially supported by Grand-in-Aid from the Ministry of Education, Culture, Sports, and Science and Technology (MEXT) of Japan, No. 15H05896, 16H05999 and 17H01110, and Bilateral Joint Research Projects of JSPS.
The International Centre for Radio Astronomy Research (ICRAR) is a Joint Venture of Curtin University and The University of Western Australia, funded by the Western Australian State government.
The MWA Phase II upgrade project was supported by Australian Research Council LIEF grant LE160100031 and the Dunlap Institute for Astronomy and Astrophysics at the University of Toronto.
This scientific work makes use of the Murchison Radio-astronomy Observatory, operated by CSIRO. We acknowledge the Wajarri Yamatji people as the traditional owners of the Observatory site. Support for the operation of the MWA is provided by the Australian Government (NCRIS), under a contract to Curtin University administered by Astronomy Australia Limited. We acknowledge the Pawsey Supercomputing Centre which is supported by the Western Australian and Australian Governments.
\end{acknowledgements}


\end{document}